\documentclass{ws-ijmpe}

\begin{document}

\markboth{G. Sood}{Production of Anti-centauro events in ...}

\catchline{}{}{}{}{}
%%%%%%%%%%%%%%%%%%%%%%%%%%%%%%%%%%%%%%%%%%%%%%%%%%%%%%%%%%%%%%%%%%%%

\title{PRODUCTION OF ANTICENTAURO EVENTS IN ULTRA-RELATIVISTIC HEAVY ION COLLISIONS}

\author{G. Sood}

\address{Physics Department, Government College (GCG -11),\\
Chandigarh, 160011, India. \\
gopika@hepphys.com}

\maketitle

\begin{history}
\received{(received date)}
\revised{(revised date)}
%\accepted{(Day Month Year)}
%\comby{(xxxxxxxxxx)}
\end{history}

\begin{abstract}
We propose a novel method for studying the production of anticentauro events in 
high energy heavy ion collisions utilizing Chebyshev expansion coefficients. These 
coefficients have proved to be very efficient in investigating the pattern of 
fluctuations in neutral pion fraction. For the anticentauro like events, the 
magnitude of first few coefficients is strongly enhanced ($\approx$3 times) as
compared to those of normal HIJING events. Various characteristics of Chebyshev
coefficients are studied in detail and the probability of formation of exotic
events is calculated from the simulated events.
 
\end{abstract}
\section{Introduction}
In ultra-relativistic heavy ion collisions the rapid expansion of collision 
debris leads to the production of vacuum states with anomalous chiral order
parameters. As a result of this, interior vacuum straighten out and radiate 
away its pionic orientation. If the deflection of vacuum orientation is in
the $\pi^{0}$ direction all the condensate radiation will be $\pi^{0}$. On
the other side, if deflection is orthogonal to the  $\pi^{0}$ direction all
the emitted pions will be charged. The former case is called ``Anti-centauro'' 
behavior and the latter is ``Centauro''. Both of these type of events are detected
in the cosmic ray experiments \cite{cosmic}. Various models \cite{models} have 
been developed to understand the characteristics of these events and numerous
signatures \cite{signal} for identification of such events are suggested.

The LHC facility \cite{lhc} at CERN will soon be operational and it would be
interesting to search for anomalous events of centauro/anticentauro nature at
LHC energies. Further, it is expected that at these high energies the production 
of exotic events will be enormous. So, it is useful to perform a theoretical 
simulation to identify the specific variables and mathematical techniques that 
may be informative in this study. The primary and widely accepted signal for
observing these events is to look for anomalous fluctuations in neutral pion
fraction (correspondingly, large fluctuations in the energy ratio, electromagnetic 
to hadronic). In this paper, we introduce a technique of Chebyshev 
coefficients \cite{cheby} to study the fluctuations in the neutral pion fraction. 

\section{Characteristics of Chebyshev polynomials}
Chebyshev polynomials are actively used in the field of mathematics for computing
the derivative/integral of a non-singular function in the finite domain and in
solving various differential equations \cite{diffeq}. In the High Energy Physics
area, this method is applied in the track reconstruction procedure \cite{trackreco},
vertex calculation and in studying the pattern of fluctuations \cite{fluc}.
The advantage of Chebyshev expansion technique is that it can be extended to 
any variable and it can be easily applied to any experimental data.     
In this paper, we have used this novel technique to study the fluctuations in the 
azimuthal distribution of neutral pions to the total number of pions produced 
in an event. The neutral pion fraction, $f(\phi)$, can be expanded in terms of
sum of Chebyshev polynomials as :

\begin{equation}
f(\phi) = \frac{C_{0}}{2} + \sum_{k=1}^{N}C_{k} T_{k}(\phi)
\end{equation}
where $T_{k}(\phi)$ is the Chebyshev polynomial of degree k.
The coefficients C$_{j}$, {\em j=0, 1, 2, $\cdots$, N-1}, of Chebyshev expansion are
defined as
\begin{equation}
 C_{j} = \frac{2}{N} \sum_{k=1}^{N} f(\phi_{k}) T_{j-1}(\phi_{k})
\end{equation}

\begin{equation}
C_{j}   = \frac{2}{N}   \sum_{k=1}^{N}  f[cos(\frac{\pi(k-\frac{1}{2})}{N})]
  cos(\frac{\pi (j-1)(k-\frac{1}{2})}{N})
\end{equation}
The coefficients of Chebyshev polynomial carry all the information on the
fluctuation of $f(\phi)$. The useful measures are provided by ranking the
coefficients, $C_{k}$, in order of the magnitude of the absolute 
values, i.e.,
\begin{equation}
\mid C_{0} \mid \hspace{0.1cm} >  \hspace{0.1cm} \mid C_{1} \mid  \hspace{0.1cm} > 
\hspace{0.1cm} \mid C_{2} \mid \hspace{0.1cm} > \hspace{0.1cm} \mid C_{3} \mid \cdots
\end{equation}
The magnitude and range of these coefficients define the strength and level
of fluctuations present in the data sample.

\begin{figure}
\centering
\includegraphics[width=3.0in]{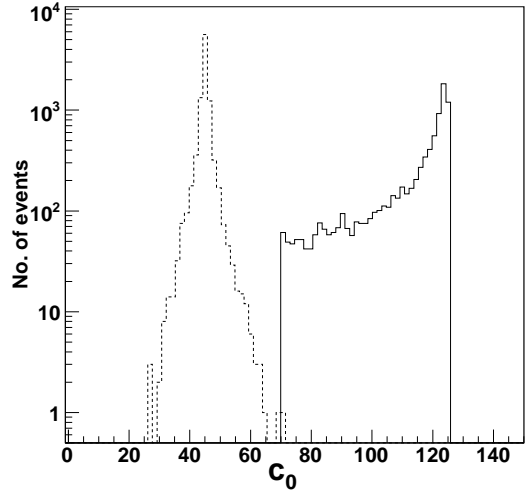}
\caption{\label{fig:epsart1}First Chebyshev coefficient, $C_{0}$, for HIJING (dashed line)
and anticentauro like events (solid line).}
\end{figure}

\section{Simulation}
The cosmic ray evidence on ``anti-centauro events''  serve as a great motivation 
for performing an exotic event search at Large Hadron Collider  (LHC). Therefore, 
for any study, in this direction, it is essential to follow the hints and inferences
provided by the cosmic ray experiments. We have taken the spectacular JACEE prototype \cite{jlord} as a reference for generating the anticentauro events 
artificially. The standard Monte Carlo program, HIJING, is used to simulate minimum
biased (0-20 fm) Pb-Pb collisions at LHC energies. Default values are taken for
all the parameters. Two ensembles of 10 K events, (i) the normal HIJING events and
(ii) the HIJING events with anticentauro type fluctuations embedded in it, are 
generated. The anticentauro event is modeled according to the ($1/2 \sqrt{f}$) 
distribution. For the present simulation, a domain of size 60$^{\circ}$ is injected
into the normal HIJING event. This is established by flipping the charged pions 
into neutral pions unless they attain a value of the neutral pion fraction, 
$f \geq 0.9$.

\section{Analysis}
We calculate the neutral pion fraction, $f$, for the generated data sets (HIJING 
as well as ``anticentauro like events'') in the $\Delta \eta - \Delta \phi$ phase space as, 
\begin{equation}
f = \frac{n_{\pi^{0}}}{n_{\pi^{0}} + n_{\pi^{\pm}}}
\end{equation}
This calculation is performed on an event-by-event basis for various azimuthal bins,
viz, $\Delta\phi$ = $2\pi/18$, $2\pi/9$, $2\pi/6$, $2\pi/4$, $2\pi/3$, $\pi$ and $2\pi$
and in the broad pseudorapidity region ($\eta$ = -10 to 10). The Chebyshev 
coefficients corresponding to the computed neutral pion fraction are determined 
according to the equation 3, as described in section II. Various characteristics of
Chebyshev polynomials are investigated in view of their ability to pick up the 
events with anomalous fluctuations in the neutral pion fraction. The details are
described below : 
 
\subsection{First order coefficients}
The first Chebyshev coefficient, $C_{0}$, for $\Delta \phi$ = $2\pi/18$, for
normal events ($f$ = 0.33) is plotted in Fig. 1. This distribution is Gaussian (dotted
line) and it is chosen as reference for studying the pattern of fluctuations in the
generated anti-centauro type events. This is an ideal choice for comparison as 
there are no hidden fluctuations present in this sample and it purely represents 
the true nature of the system. In the same figure, we have plotted the first order
coefficient ($C_{0}$), for the anticentauro like events (solid line). The absolute
value of the magnitude of this coefficient is large as compared to that of 
normal events. Similar behavior is observed for other order coefficients, i.e., 
for the exotic events the value of coefficients is significantly enhanced as 
compared to the normal events. Table 1. describes the detailed results of the 
variation of absolute magnitude of first order coefficient, $C_{0}$, with azimuth,
for the top 5\% central events. The mean value of the coefficient, $C_{0}$, for
anticenturo like events is approximately thrice the mean value of normal HIJING events.
 It is observed that with increase in size of domain 
the magnitude of coefficients decreases. This behavior is seen for both HIJING 
and anticentauro like events but the magnitude of the coefficients for exotic
events is much more than normal events for all the azimuthal bins. 

\begin{table}[pt]
\tbl{Mean value of the first coefficient, $C_{0}$, for the exotic
and normal HIJING events.}
{\begin{tabular}{ccc} \toprule
 &\multicolumn{2}{c}{Mean $C_{0}$} \\
 Azimuthal size&Anticentauro like event&HIJING event  \\ \colrule
$2\pi/18$ \hphantom{00} & \hphantom{0}$124.8$ & \hphantom{0}$45.0$ \\
$2\pi/9$ \hphantom{00} & \hphantom{0}$63.2$ & \hphantom{0}$23.0$ \\
$2\pi/6$ \hphantom{00} & \hphantom{0}$42.7$ & \hphantom{0}$15.4$ \\
$2\pi/4$ \hphantom{00} & \hphantom{0}$28.5$ & \hphantom{0}$10.2$ \\
$2\pi/3$ \hphantom{00} & \hphantom{0}$21.4$ & \hphantom{0}$7.7$ \\
$\pi$ \hphantom{00} & \hphantom{0}$14.3$ & \hphantom{0}$5.2$ \\
$2\pi$ \hphantom{00} & \hphantom{0}$7.1$ & \hphantom{0}$2.6$\\ \botrule
\end{tabular}}
\end{table}

\subsection{Higher order coefficients}
The coefficients of higher degree Chebyhshev polynomial, C$_{k}$ for 0$\le$ k $\le$ N-1
contain all the information of fluctuations. In our study, we have calculated 100
coefficients, $C_{0}$ to $C_{99}$. All these coefficients, $C_{k}$ (k = 0, 99) are
pure Gaussian (for normal HIJING events), however, the absolute values of their 
magnitude decreases very rapidly. Fig. 2 shows the distribution of some higher order
coefficients, $C_{30},C_{40},C_{50}$ and $C_{90}$. It is observed that higher order 
coefficients gradually attain the null value. Therefore, the Chebyshev expansion
series can be truncated at some order $m < N$ and the extreme higher order
coefficients are neglected. It is important to note that the absolute value of the
magnitude of higher order coefficients for the ``anticentauro like events'' is still
much higher than the corresponding HIJING events as observed for the first order 
coefficient.   

\subsection{Correlation of Chebyshev coefficients}
The magnitude and range of the Chebyshev coefficients is very useful in predicting 
whether or not the fluctuations in the neutral pion fraction are of non-statistical
nature. A more critical criteria can be evolved by considering the two coefficients 
simultaneously. For this, it is necessary to study the correlation between various
coefficients. Fig 3. shows the relationship between the two consecutive coefficients 
$C_{1}$ and $C_{2}$. It is seen that the coefficients are correlated i.e., with 
increase in $C_{1}$, the coefficient $C_{2}$ also increases. Similar behavior is
observed for other coefficients. The corresponding magnitude of coefficients for  
HIJING events is very less.

\begin{figure}[h]
  \hfill
  \begin{minipage}[t]{.45\textwidth}
    \begin{center}  
      \epsfig{file=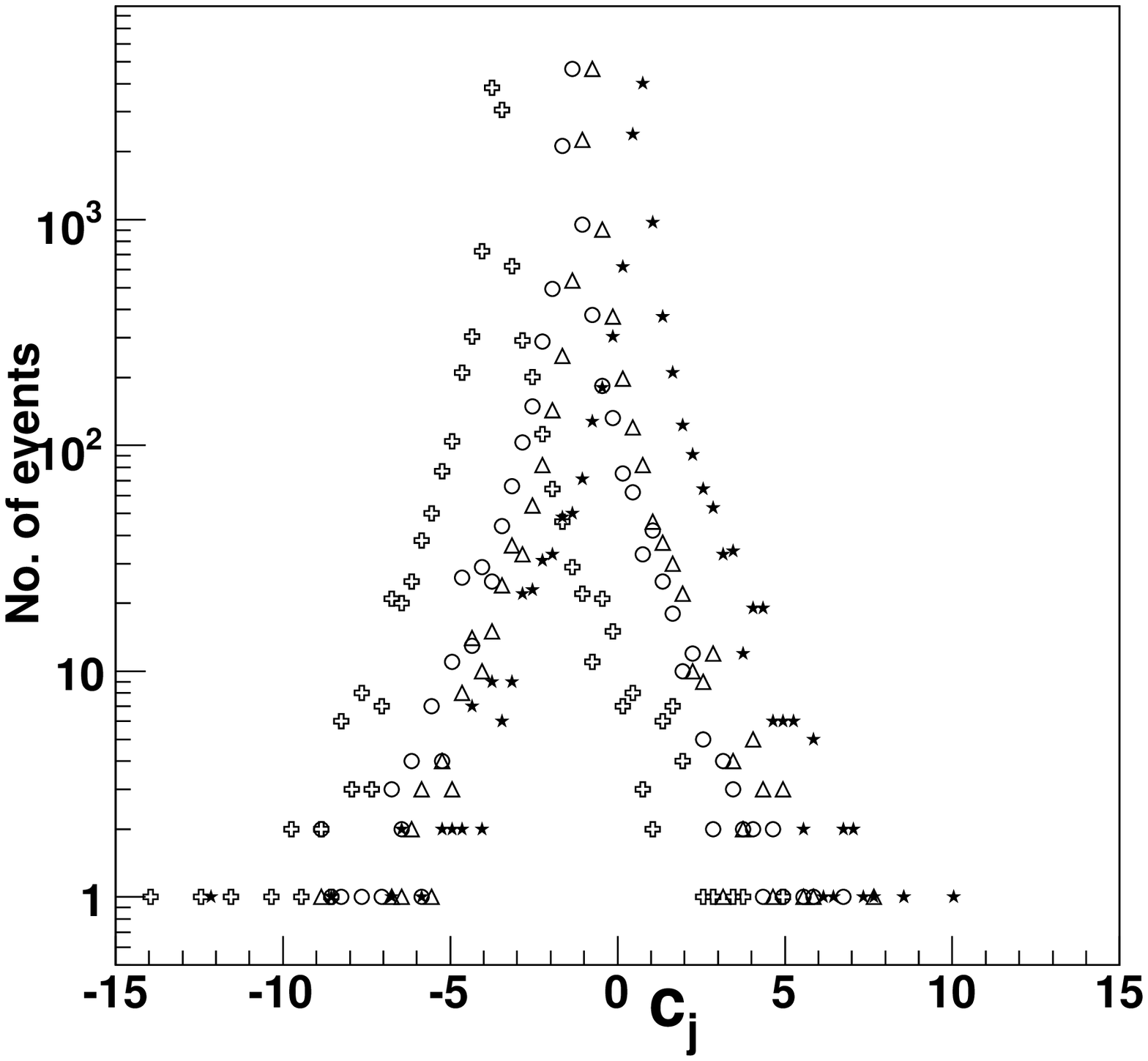, scale=0.32}
      \caption{The distribution of higher order coefficients, $C_{30}$, $C_{40}$,
 $C_{50}$ and $C_{90}$ for the minimum biased HIJING events.}
      \label{fig-tc}
    \end{center}
  \end{minipage}
  \hfill
  \begin{minipage}[t]{.45\textwidth}
    \begin{center}  
      \epsfig{file=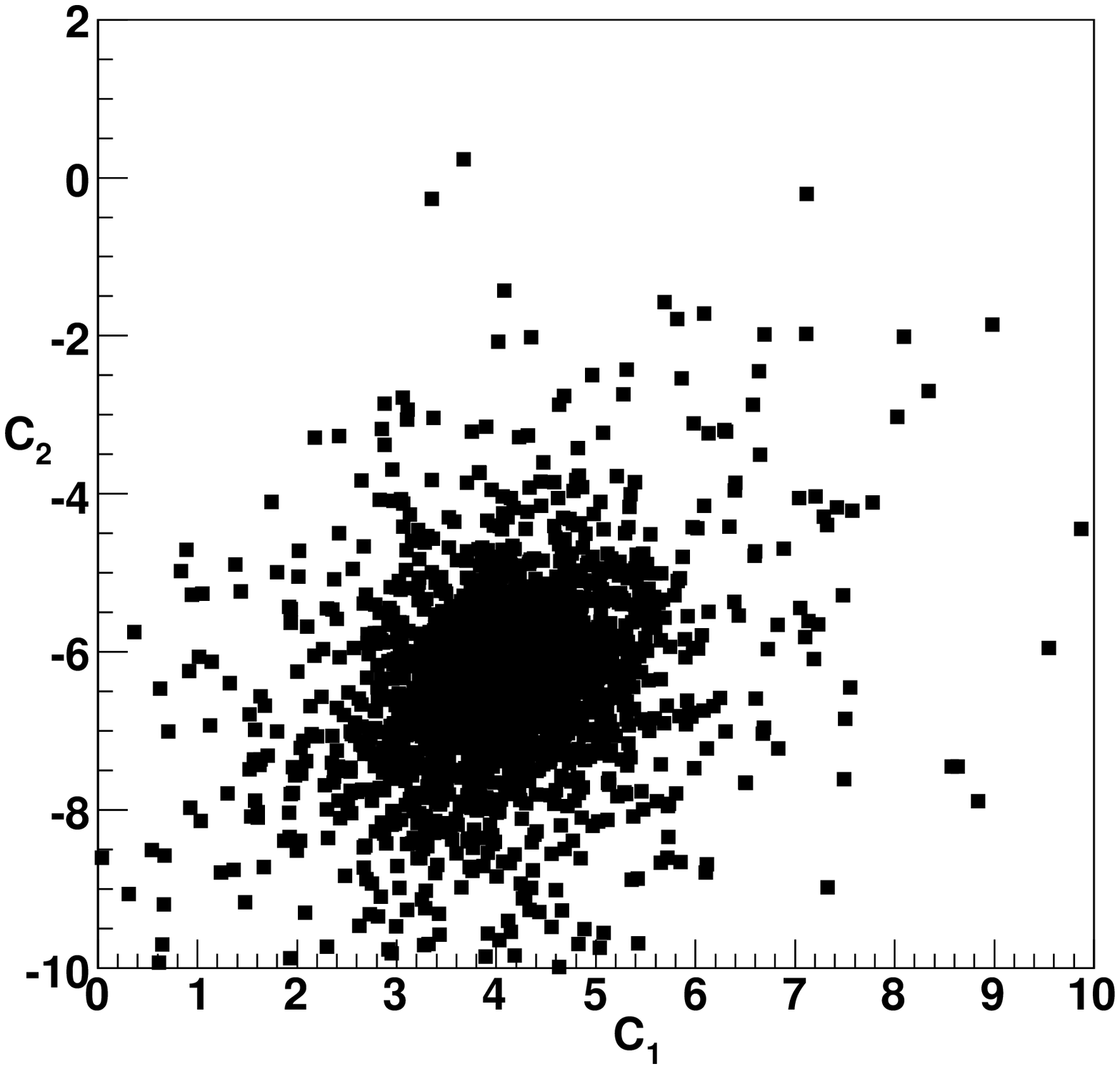, scale=0.32}
      \caption{Correlation plot for the coefficients, $C_{1}$ and $C_{2}$.}
      \label{figtc}
    \end{center}
  \end{minipage}
  \hfill
\end{figure}

\subsection{Probability function: An estimate from simulation}
A more quantitative way to analyze fluctuations is to estimate the production of 
anticentauro events from simulation. For this, we define a probability function P,
which is the fraction of events which have the magnitude of first order coefficient,
$C_{0}$, greater than the threshold set by the sum of mean first order coefficient,
obtained from HIJING events and five times the dispersion ($<C_{0}> + 5\sigma$).
For all the anticentauro events, the absolute value of the magnitude of first
order coefficient is much greater than the set threshold but for the HIJING 
events we do not observe any event which satisfies the above criteria. As an
example, for the azimuthal bin, $\Delta \phi$ = $2\pi/6$, all the artificially
generated anticentauro events have magnitude of the first order coefficient 
($C_{0}$ = 42.7) which is much higher than the threshold value (17.21). The 
corresponding statistical probability obtained from HIJING events is
\begin{equation}
P(C_{0} > 17.21) = 0
\end{equation}
For the semi-central events the probability is
\begin{equation}
P(C_{0} > 16.61) = 2.2 {\tt x} 10^{-3}
\end{equation}
Similar behavior is observed for the other azimuthal bins. Thus, for the generated
exotic events, high value of neutral pion fraction in the $\eta$ -$\phi$ phase space 
is clearly exhibited by high magnitude of the Chebyshev coefficient whereas 
statistically the probability of formation of exotic event is either 0 or very small. 

\section{Summary}
In this study, we have described a method of Chebyshev polynomials to study the 
fluctuations in the neutral pion fraction of artificially generated anticentauro
events for Pb+Pb collisions at LHC energies. This method is proved to be very
efficient to measure the strength of fluctuations. Our analysis shows that the 
magnitude of higher range first order coefficients for exotic events is 
approximately 3 times greater than those of normal events. This trend is seen for
other higher range coefficients. The detailed analysis is carried out by estimating 
the probability of formation of exotic events from the normal events.

\end{document}